# Photocatalytic methanol dehydrogenation promoted synergistically by atomically dispersed Pd and clustered Pd

Zhuyan Gao,[1,2,5] Tiziano Montini,[3,5] Junju Mu,[1] Nengchao Luo,[1,*] Emiliano Fonda,[4] Paolo Fornasiero,[3,*] and Feng Wang[1,2,*]

[1] State Key Laboratory of Catalysis, Dalian National Laboratory for Clean Energy, Dalian Institute of Chemical Physics, Chinese Academy of Sciences, Dalian, 116023, China.

[2] University of Chinese Academy of Sciences, Beijing, 100049, China.

[3] Department of Chemical and Pharmaceutical Sciences, Center for Energy, Environment and Transport Giacomo Ciamiciam, INSTM Trieste Research Unit and ICCOM-CNR Trieste Research Unit, University of Trieste, Via Licio Giorgieri 1, 34127 Trieste, Italy

[4] Synchrotron SOLEIL, L'Orme des Merisiers, Saint Aubin BP48, 91192 Gif sur Yvette CEDEX, France

[5] These authors contributed equally: Zhuyan Gao, Tiziano Montini.

* Corresponding Author.
*N. L.: e-mail, ncluo@dicp.ac.cn.
*P. F.: e-mail, pfornasiero@units.it
*F. W.: e-mail, wangfeng@dicp.ac.cn

**Graphical Abstract**

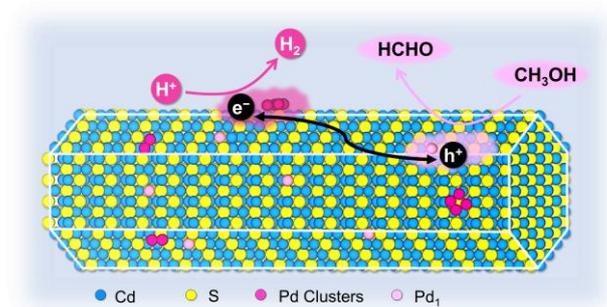

**Abstract**

Supported metal in the form of single atoms, clusters, and particles can individually or jointly affect the activity of supported heterogeneous catalysts. While the individual contribution of supported metal to the overall activity of supported photocatalysts has been identified, the joint activity of mixed metal species is overlooked because of their different photoelectric properties. Here, atomically dispersed Pd ($Pd_1$) and Pd clusters are loaded onto CdS, serving as oxidation and reduction sites, respectively, for methanol dehydrogenation. The $Pd_1$ substitutes $Cd^{2+}$, forming hole-trapping states for methanol oxidation and assisting the dispersion of photo-deposited Pd clusters. Therefore, methanol dehydrogenation on CdS with supported $Pd_1$ and Pd clusters exhibits a highest turnover frequency of 1.14 $s^{-1}$ based on Pd content, and affords $H_2$ and HCHO with a similar apparent quantum yield of 87 ± 1% at 452 nm under optimized reaction conditions. This work highlights the synergistic catalysis of supported metal for improved photocatalytic activity.

# Introduction

Dehydrogenation of liquid organic $H_2$ carriers is a promising way for on-site $H_2$ production to feed the proton-exchange membrane fuel cell.[1] Methanol tallies with the requirements for the liquid organic $H_2$ carrier since it is easily available from coal,[2] biomass,[3] and $CO_2$,[4-5] and has a high mass proportion of hydrogen (12.5 wt%). Recent advances in systematic solar energy utilization have identified methanol as the "liquid sunshine", which is regarded as the most promising liquid to store solar energy.[6-7] Releasing $H_2$ from methanol is consequently significant to fulfilling the cycle for the $H_2$ economy but suffers from thermodynamic limitations (methanol dehydrogenation, $\Delta_r G^\ominus_m = 59.8$ kJ mol$^{-1}$).[8] Photocatalysis can conquer the challenge by capitalizing on solar energy and a highly efficient and visible-light-responsive photocatalyst.[9-11] Cu single atoms supported on $TiO_2$ were evidenced to efficiently dehydrogenate methanol with an apparent quantum yield (AQY) of 56% for $H_2$ production at 365 nm.[12] Besides, combining Cu single atoms with Pt nanodots on $TiO_2$, the AQY at 365 nm for methanol dehydrogenation was increased from 56.2% at room temperature to 99.2% at 70 °C,[13] representing the benchmark of photocatalytic methanol dehydrogenation. This pioneering work highlights the synergy of single atoms and clusters. However, the UV-light absorption feature and extra heating of the photocatalytic system would raise additional costs on investments in equipment.

The size of supported metals critically affects the photocatalytic reactions because of their distinct chemical environment and quantum confinement effects.[14-16] Generally, noble metal nanoclusters or particles are conceived as the active sites for $H_2$ evolution reaction,[17-19] while low-work-function metals can tune the electronic properties of the semiconductor support, thus fostering C−H bond scission.[20] The catalytic effects of single atoms are more complex as they can be the catalytic sites or form defect levels in the bandgap of the semiconductor support. For example, single atoms such as Pt on $TiO_2$ were evidenced to catalyze $H_2$ evolution while Ag single atoms in CdS were identified to level down the conduction band potential, thus promoting the reductive cleavage of C−O bonds in lignin.[21-22] Cu single atoms have multiple effects according to the valence states, such as acting as the oxidation site for methanol oxidation, promoting $H_2$ evolution, and promoting charge separation.[12-13, 23]

Herein, capitalizing on the complexity of supported metal species, we demonstrated that atomically dispersed ($Pd_1$) and clustered Pd on a visible-light-responsive CdS semiconductor (absorbs light above 520 nm)[24] can synergistically catalyze methanol dehydrogenation (Figure 1). The $Pd_1$ was loaded on CdS by ion exchange with lattice $Cd^{2+}$ in an aqueous solution, originating defect energy levels capable of trapping photogenerated holes for methanol oxidation. Clustered Pd was loaded on the Cd(Pd)S by second-step photo-deposition in methanol solution, with the sizes restrained thanks to the pre-loaded $Pd_1$. Methanol was dehydrogenated on Pd/Cd(Pd)S with a highest turnover frequency (TOF) of 1.14 s$^{-1}$, producing $H_2$ and HCHO with similar AQYs of 87 ± 1% at 452 nm, which surpasses the benchmark photocatalysts at room temperature. Notably, over 20% of methanol has been used for HCHO production industrially,[25] which uses silver-based and iron-molybdenum-based catalysts at high temperatures (> 300 °C).[26] Such synergy between single atoms and clusters of the same metal thus highlights the significance of capitalizing on the complexity of supported metals for promising coproduction of $H_2$ and value-added chemicals.

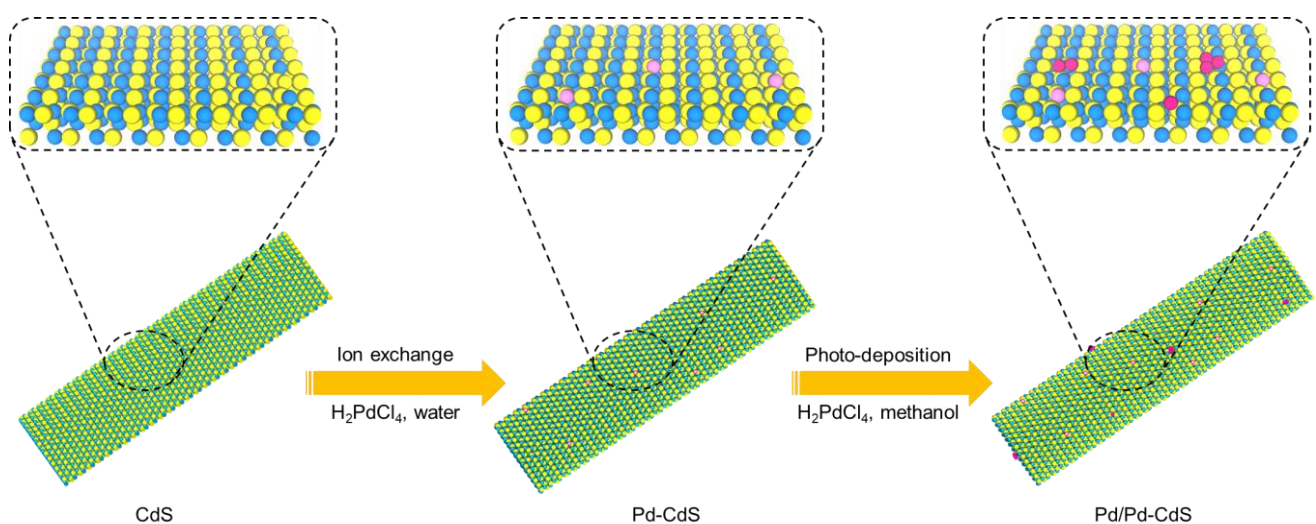

**Figure 1**. Schematic illustration of the preparation procedures for loading atomically dispersed Pd ($Pd_1$) and clustered Pd on CdS photocatalyst. $Pd_1$ is loaded on CdS surface by ion exchange with surface $Cd^{2+}$ in aqueous solution to form Cd(Pd)S. Pd clusters are loaded on Cd(Pd)S by photo-deposition in methanol solution to form Pd/Cd(Pd)S. The pre-loaded $Pd_1$ can restrain the sizes of Pd to small clusters. Cd (blue), S (yellow), atomically dispersed Pd (pink), clustered Pd (rose red).

## Results and discussion

### Characterization of prepared catalysts

Pd-exchanged CdS (Cd(Pd)S) was prepared by dispersing CdS into the aqueous solution of $H_2PdCl_4$, as the solubility product constant (p$K_{sp}$) of PdS (57.7) is greater than that of CdS (26.1).[8, 27] The ion exchange between the $Pd^{2+}$ of $H_2PdCl_4$ and $Cd^{2+}$ on CdS surface was verified by inductively coupled plasma atomic emission spectroscopy (ICP-AES). The aqueous suspension containing CdS and 0.19 mM of $H_2PdCl_4$ was stirred in dark for 3 h. After filtration of the aqueous suspension, analysis of the filtrate by ICP-AES revealed that Pd disappeared while 0.26 mM of $Cd^{2+}$ was found in the filtrate, consistent with the occurrence of the ion exchange between $Pd^{2+}$ and $Cd^{2+}$ on CdS. In comparison, after filtration of the methanolic suspension containing CdS and $H_2PdCl_4$, Pd disappeared in the filtrate while $Cd^{2+}$ had a concentration of 0.06 mM. The observation of $Cd^{2+}$ was due to the acidolysis of CdS while the adsorption of $PdCl_4^{2-}$ on CdS can rationalize the inability to detect $Pd^{2+}$ in the filtrate.

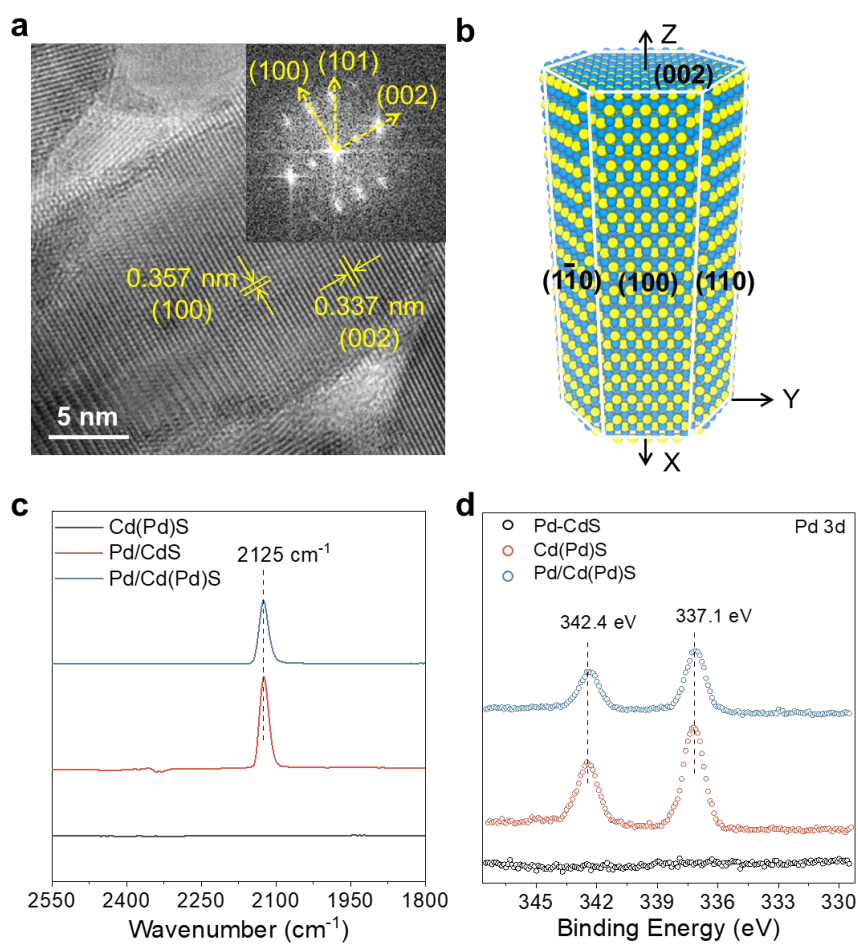

**Figure 2. Characterization of the Pd-modified CdS photocatalysts.** (a) High-resolution TEM image of the Pd/Cd(Pd)S catalyst. The inset shows the selected area electron diffraction pattern. (b) Crystal structure of the Pd/Cd(Pd)S built by referring to the XRD and TEM results. Cd (blue), S (yellow). (c) CO-adsorption diffuse reflectance infrared Fourier transform spectroscopy. (d) X-ray photoelectron spectroscopy measured at Pd 3*d* regions.

After synthesizing Cd(Pd)S in aqueous suspension, additional Pd was loaded on the Cd(Pd)S by photo-deposition. The Pd contents of Cd(Pd)S and Pd/Cd(Pd)S were determined to be 0.1 and 1.3 at% (relative to Cd), respectively, by ICP-AES. X-ray diffraction (XRD) patterns of Cd(Pd)S and Pd/Cd(Pd)S showed diffraction peaks well attributed to hexagonal wurtzite structure (Figure S1a). The diffraction peak at 26.5º was sharper than those at 24.8º and 28.2º, indicating the prepared samples formed nanorods according to the analytic results of mean crystallite size calculated by applying the Scherrer equation to the above-mentioned reflections (Table S1). Diffraction peaks ascribed to Pd nanoparticles or PdS were not observed, indicating that Pd was highly dispersed on CdS. In the transmission electron microscopy (TEM) image (Figure S1b), the CdS was in the form of nanorods with a length of about several hundred nanometers and a diameter between 20-30 nm. The observed lattice fringes with distances of 0.357 and 0.337 nm in the high-resolution TEM were attributed to the (100) and (002) crystal planes of hexagonal CdS, respectively (Figure 2a). The pattern of selected area electron diffraction (SAED) suggested that CdS nanorods existed as hexagonal prism composed of the (002) plane as the end plane and six side planes equivalent to {100} planes (Figures 2a and 2b). Because the atomic numbers of Pd and Cd are very close, Pd species were not accurately found in the TEM images.

The oxidation state of Pd was inferred by diffuse reflectance infrared Fourier transform (DRIFT) spectroscopy. CO was used as the probe molecule in acquiring the DRIFT. Samples for analysis were pretreated at 120 °C in Ar atmosphere and cooled to 30 °C before recording the spectra. Signals ascribed to CO adsorption were not observed for Cd(Pd)S samples (Figure 2c), which may be ascribed to the coordination of residual $Cl^-$ to Pd or the low Pd content (0.1 at%). Signals of CO appeared at a wavenumber of 2125 $cm^{-1}$ for the Pd/CdS and Pd/Cd(Pd)S samples. The signal attributed to CO adsorbed on $Pd^{2+}$ is usually located at 2215-2145 $cm^{-1}$ in DRIFT spectra for metal oxide supported $Pd^{2+}$.[28] Compared with metal oxides, the chemical bond between S and oxidized Pd of CdS is more covalent, so the signal tends to shift towards low wavenumbers for CO adsorbed on $Pd^{2+}$ loaded on CdS. Consistent with this conclusion, some literatures have reported that CO adsorbed on $Ni^{2+}$ supported on $Al_2O_3$ and sulfurized $Al_2O_3$ were located at 2190 and 2080 $cm^{-1}$, respectively.[29-30] Therefore, the signal at 2125 $cm^{-1}$ in Figure 2c is ascribed to Pd with an oxidation state of +2, suggesting that the photo-deposited Pd may form tiny Pd clusters or single atoms. The oxidation state of Pd was further confirmed by X-ray photoelectron spectroscopy (XPS). Pd 3d XPS signal was not observed for the Cd(Pd)S sample, due to the low Pd content (0.1 at%). As for the Pd/CdS and Pd/Cd(Pd)S samples, Pd 3d XPS signal showed binding energy of 342.4 and 337.1 eV (Figure 2d), attributed to Pd $3d_{3/2}$ and Pd $3d_{5/2}$ spectra, respectively, with a Pd oxidation state of +2.[31-32]

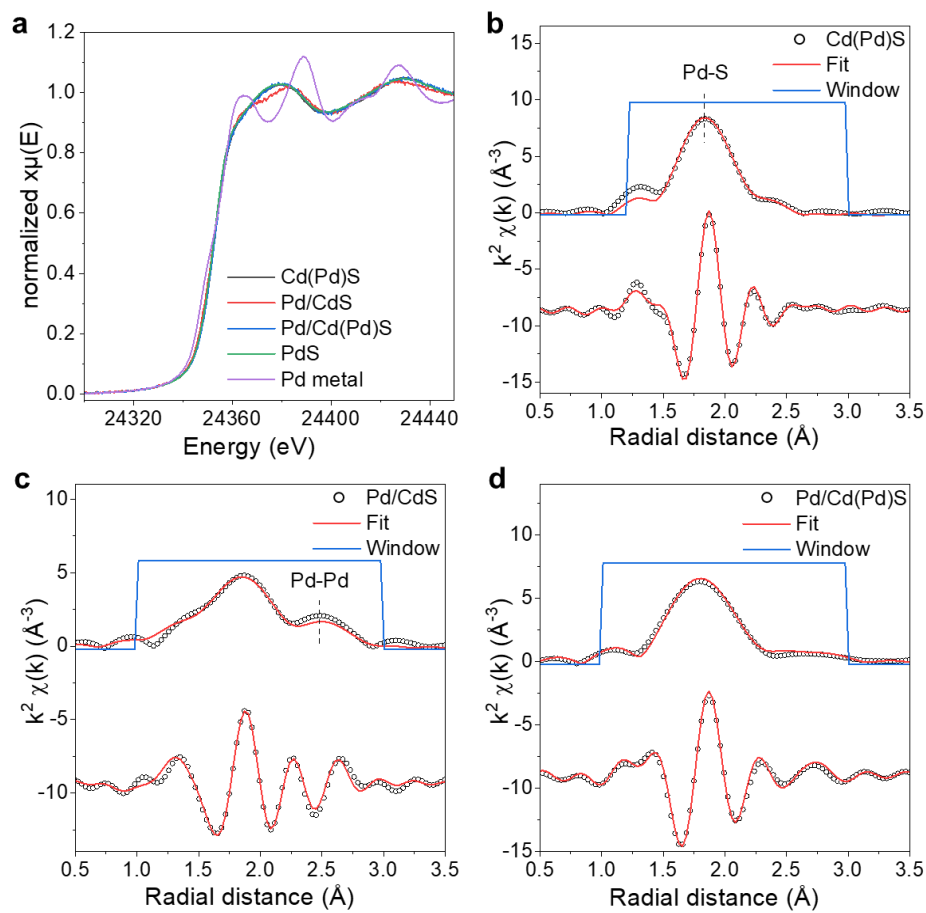

**Figure 3.** Characterization of irradiated Cd(Pd)S, Pd/CdS, and Pd/Cd(Pd)S catalysts by extended X-ray absorption fine structure (EXAFS). (a) Pd K-edge XANES spectra of Cd(Pd)S, Pd/CdS, and Pd/Cd(Pd)S. (b) Fits of FT of $k^2\chi(k)$ of Cd(Pd)S EXAFS signals. (c) Fits of FT of $k^2\chi(k)$ of Pd/CdS EXAFS signals. (d) Fits of FT of $k^2\chi(k)$ of Pd/Cd(Pd)S EXAFS signals.

X-ray absorption near-edge structure (XANES) and extended X-ray absorption fine structure (EXAFS) spectroscopy were performed to analyze the change of Pd oxidation states and coordination environments during ion exchange and photo-deposition (Figure 3a). It must be underlined that the scattering power of S and Cl cannot be distinguished when fitting EXAFS spectra. A Pd−S path has been employed for all the samples containing the CdS photoactive material, even though the presence of residual Cl atoms cannot be excluded. Table S2 presents the fit of EXAFS signals of relevant standards (Pd metal, PdS, and $K_2PdCl_4$) for clarity.

After dissolution in $H_2O$, the EXAFS signal of $H_2PdCl_4$ can be reconstructed by the contribution of Pd−Cl and Pd−O paths (Figure S2a and Table S3), indicating that coordinated water molecules can easily replace $Cl^-$.[33-34] The low concentration of $Cl^-$ allowed the formation of hydration ions which more easily absorb on the CdS surface than $PdCl_4^{2-}$.[35] The formation of hydrated ions is also beneficial to ion exchange with $Cd^{2+}$. After interacting with CdS, the XANES spectrum of the resulting ion-exchanged Cd(Pd)S in an aqueous solution is very close to that of PdS (Figure S3a). Further analysis of the EXAFS signal (Figure S3b) showed that the fitting is acceptable including the Pd−S path with a coordination number (CN) of 3.9 and a Pd−Pd path with CN of 0.5 (Table S3), in which the distances were in reasonable agreement with the values obtained in PdS and metal Pd (Table S3). This result suggests that, after the interaction of CdS with $H_2PdCl_4$ in an aqueous solution, $Pd^{2+}$ mainly replaces $Cd^{2+}$ in the CdS lattice and reasonably forms a small portion of Pd clusters, strongly interacting with S atoms of the support. After irradiation (Figure 3b, Table S3), only a marginal decrease of the CN of the Pd−S path was observed, probably corresponding to the removal of $Cl^-$,[36-37] indicating that $Pd^{2+}$ is hardly reduced when hosted in the CdS structure.

The EXAFS signal of $H_2PdCl_4$ in methanolic solution was similar to that in aqueous solution (Figure S2b and Table S3) but showed significant differences after interacting with CdS. After adsorption of $H_2PdCl_4$ in the methanolic suspension of CdS and Cd(Pd)S, ICP-AES analysis of the filtrate indicates that almost all of the $H_2PdCl_4$ adsorbs on the catalysts in the dark. The XANES spectrum of the sample obtained by adsorption of $H_2PdCl_4$ on CdS in methanolic solution can be fitted as a linear combination of XANES spectra of PdS and $H_2PdCl_4$ in methanolic solution (Figure S4a) while the corresponding EXAFS signal can be fitted assuming (Figure S3c), beside the main contribution of Pd−S path (CN = 3.0), the presence of a Pd−O path with a CN of 1.3 (Table S3). These results approved the formation of Pd species adsorbed on CdS surface through S atoms but not enter the CdS lattice. After illumination, the EXAFS signal of Pd/CdS is significantly modified (Figure 3c): CN for Pd−O and Pd−S were reduced to 0.8 and 2.0, respectively, and a Pd−Pd path with CN = 0.7 appeared in the fitting result. The linear combination fitting of the XANES spectrum of this sample showed that 17% of total Pd had been reduced to the zero-valent state (Figure S4b): assuming that the Pd−Pd distance originates only from the zero-valent fraction of Pd, the average CN in metallic part results to be close to 4.1, a value correspondent to tiny Pd nanoparticles (NPs). The average Pd−Pd distance was 0.277 nm, which was close to the value observed for metallic Pd (0.274 nm). The above results indicated that the Pd species were firstly adsorbed on CdS surface during photo-deposition process,[38] and then partially reduced to small clusters by photogenerated electrons.

When the support was changed to Cd(Pd)S, the XANES spectrum after interaction with $H_2PdCl_4$

in methanolic solution still resembled that obtained after adsorption on CdS, while the EXAFS spectrum can be fitted including Pd−O, Pd−S, and Pd−Pd paths (Figure S3d). This situation revealed that, after equilibration in the dark, both $Pd^{2+}$ exchanged in CdS lattice (from Cd(Pd)S) and Pd species adsorbed on the surface (from $H_2PdCl_4$ in methanolic solution) were present. After irradiation, although the XANES spectrum was not significantly affected, the EXAFS signal showed important modifications (Figure 3d): the Pd−O path disappeared while the contribution of Pd−Pd increased. The former agreed with the reduction of CN in Pd−O observed for Pd/CdS after irradiation. For Pd−Pd path, the increase in CN was accompanied by an elongation of the average distance (0.285 nm), longer than the distance between Pd atoms in metallic Pd crystals and most complexes.[39-40] According to literature reports, the elongation of atomic distances usually occurs in some small metal clusters.[41] The overall results suggest that, in the Pd/Cd(Pd)S photocatalyst, Pd species were distributed into two components: $Pd^{2+}$ ions as substituent into CdS lattice as atomically dispersed Pd ($Pd_1$) and highly dispersed Pd clusters adsorbed on the surface and strongly interacting with surface S.

Considering the similarity in atomic number between Pd and Cd, the possibility of Pd−Cd interaction has been considered by trying to fit the EXAFS signal of the Pd/Cd(Pd)S sample with a Pd−Cd scattering path. For this reason, the EXAFS signal reported in Figure 3d was fitted using a second model (Model 2) including combination of Pd−S and Pd−Cd and with a third model including combination of Pd−S, Pd−Pd and Pd−Cd (Model 3). The graphical results of the fits were presented in Figure S5 and the structural parameters extracted from these models are summarized in Table S4. Replacing Pd with Cd (Model 2 in Table S4) leads to a structural model very similar to the first one (CN, distances and errors are very similar to those obtained considering the Pd−Pd scattering path). Despite this, the statistical significance of the model comprising Pd−Cd path is lower, as indicated by the higher value of the reduced $\chi^2$. Moreover, when both Pd−Pd and Pd−Cd paths are considered (Model 3), the reduced $\chi^2$ increased a lot (mainly because of the increased number of parameters introduced in the fit) and meaningless values have been obtained for the CN. This is because Pd and Cd are exchangeable from the point of view of EXAFS signal due to their very close atomic numbers. Consequently, the fitting parameters of Pd−Pd and Pd−Cd paths are strongly correlated. In conclusion, the model assuming $Pd^{2+}$ substituting $Cd^{2+}$ in the CdS in combination with very tiny, highly dispersed Pd clusters seem to be the more realistic one, finally excluding the reduction of $Cd^{2+}$ with formation of bimetallic clusters. This agrees with the differences in the standard reduction potentials of the two ions ($E^0(Pd^{2+}/Pd) = + 0.90$ V; $E^0(Cd^{2+}/Cd) = - 0.40$ V).

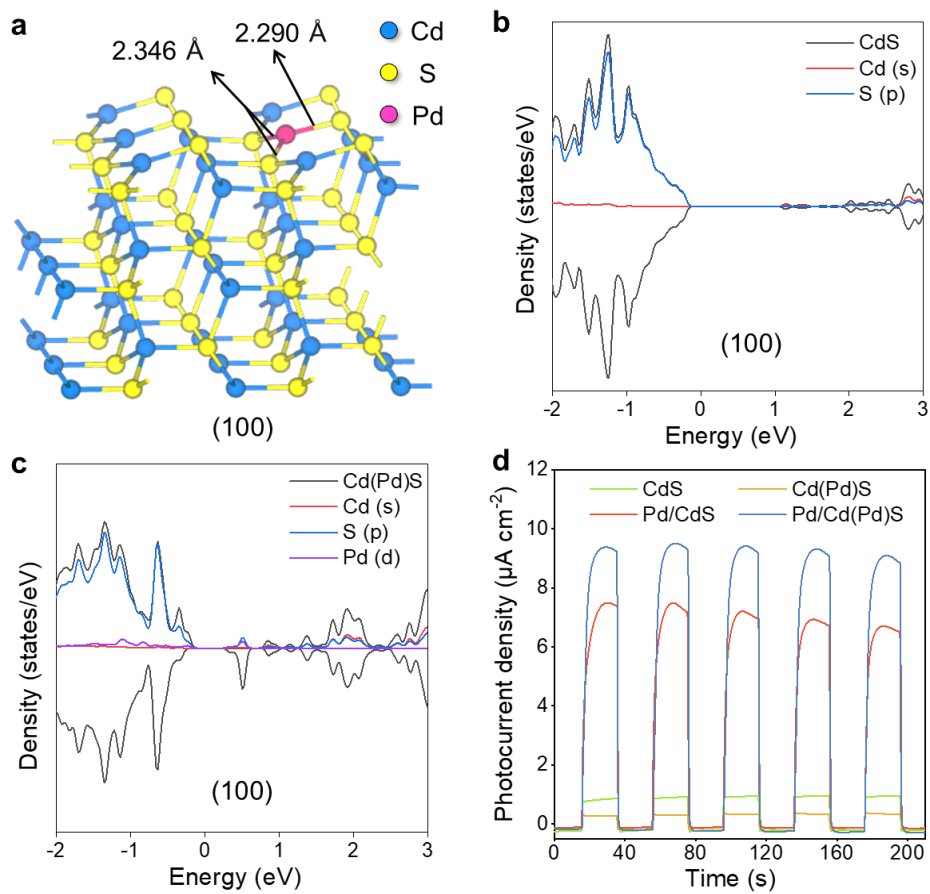

**Figure 4.** Structure model and calculated density of states (DOS) of Cd(Pd)S along with the photocurrents. (a) Optimized Cd(Pd)S structure. (b) DOS of CdS using a CdS (100) $p(3\times3\times3)$ surface model. (c) DOS of Cd(Pd)S using a Cd(Pd)S (100) $p(3\times3\times3)$ surface model. (d) Transient photocurrent density of the photocatalysts measured in methanol.

**Effects of Pd$_1$ and Pd clusters on the properties of CdS**

The influence of Pd$_1$ and Pd clusters on the band structure and photo-physical properties of CdS was studied by DFT calculations and transient photocurrents. Before studying the effect of Pd$_1$ on the electronic structure of CdS, the exact location, including (001), (00−1), and (100) crystal planes, were considered for Pd$_1$. The (100) crystal plane of CdS was identified as the most plausible plane that Pd$^{2+}$ locates (Figure 4a), as the lowest substitutional energy of −0.90 eV was obtained for the exchange of Pd$^{2+}$ and Cd$^{2+}$. In the optimized structure, Pd$^{2+}$ coordinated with three S$^{2-}$ on the surface with Pd−S distances of 2.290, 2.346, and 2.346 Å, respectively. The average Pd−S distance was very close to the average Pd−S distance inferred from EXAFS fitting (2.32 Å), approving that Pd substitutes Cd$^{2+}$ on the (100) crystal plane. The influence of $Pd_1$ on the band structure of CdS was evaluated by comparing the density of states (DOS) of CdS and Cd(Pd)S. The conduction band minimum (CBM) and valence band maximum (VBM) of pristine CdS were constituted by Cd 5s and S 3p, respectively (Figure 4b), consistent with the literature result.[42] Substitution of $Cd^{2+}$ by $Pd^{2+}$ originated mid-gap states comprised mainly of Pd 4d and S 3p, which lied close to the VBM of CdS (Figure 4c). The mid-gap states are expected to trap photogenerated holes, indicating that the $Pd_1$ is the active site for methanol oxidation. Consistently with this, the band gap of Cd(Pd)S is similar with pristine CdS (2.4 eV) according to the UV-vis diffuse reflectance spectra (UV-vis DRS) of Cd(Pd)S and CdS (Figures S6a and S6b). Moreover, an absorption tail arises for Cd(Pd)S, consistent with the mid-gap states shown in the DOS of Cd(Pd)S.

The $Pd_1$ and Pd clusters have been found to promote charge separation. Compared with pristine CdS, photoluminescence spectrum (PL) of Cd(Pd)S showed weaker emission than that of pristine CdS (Figure S7), indicating that the $Pd_1$ suppresses charge recombination. This result can be understood as $Pd_1$ can trap photogenerated holes, restraining the recombination with electrons. Deposition of Pd clusters on Cd(Pd)S further decreases the PL emission intensity, which can be rationalized by electron transfer to Pd clusters, thereby restraining charge recombination. This hypothesis could be supported by the results of transient photocurrents. Pd/Cd(Pd)S originated a weak photocurrent density in acetonitrile. The photocurrent density was enhanced when methanol consumed photogenerated holes (Figure S8), indicating the positive photocurrent is derived from photogenerated electrons that migrate to the electrode.[43] Compared with CdS, the photocurrent density of Cd(Pd)S was smaller (Figure 4d), which is rational as photogenerated holes of Cd(Pd)S are readily transferred to the surface and recombines with electrons on the surface, thus decreasing the amount of electrons transferred to the electrode. This result was consistent with the observed mid-gap states of Cd(Pd)S that enabled hole trapping. In contrast, the photocurrent density of Pd/CdS was larger than that of CdS, indicating photo-deposited Pd predominantly fostered electron transfer to CdS surface. The copresence of $Pd_1$ and Pd clusters well promote hole and electron migration, respectively, so the highest photocurrent density was derived.

**Photocatalytic anaerobic dehydrogenation of methanol**

The photocatalytic activity of the Pd/Cd(Pd)S catalyst for methanol dehydrogenation was compared with Cd(Pd)S and Pd/CdS under the optimized conditions (Figures S9 and S10). Photocatalytic methanol dehydrogenation on CdS produced $H_2$ with productivity of 0.8 mmol $g_{catal.}^{-1}$ in 8 h. By capturing the radical intermediates by 1,1-diphenylethylene, it was possible to prove the formation of hydroxymethyl radical (•$CH_2OH$) with a formation rate of 0.55 mmol $g_{catal.}^{-1}$ $h^{-1}$ (Figures S11a and S11b), suggesting the occurrence of the back reaction between •H and •$CH_2OH$ and justifying the fact that the net reaction was slow.[44] The productivity of $H_2$ dramatically increased to 216.3 mmol $g_{catal.}^{-1}$ when Pd was loaded onto CdS by photo-deposition (Pd/CdS).

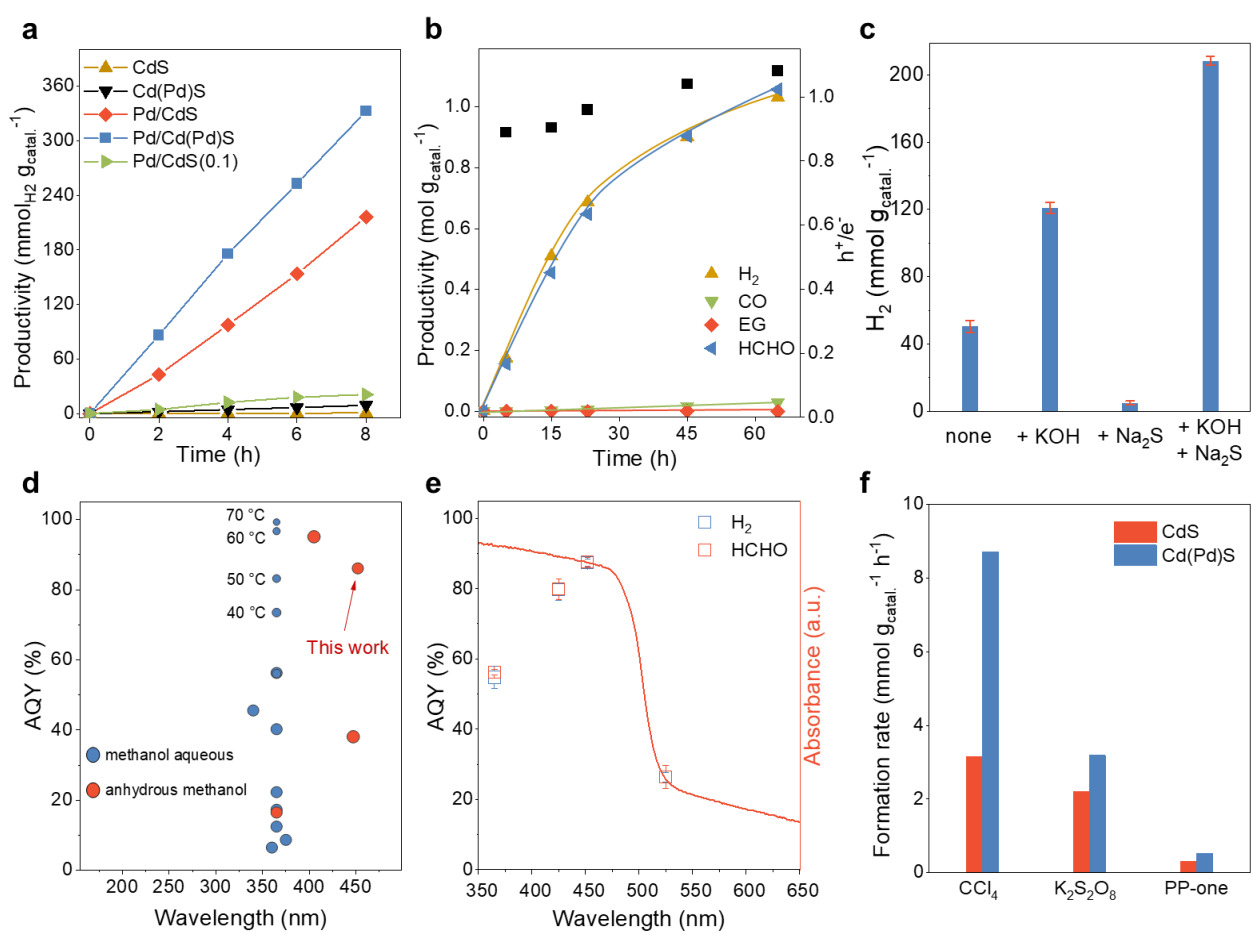

**Figure 5.** Photocatalytic anaerobic dehydrogenation of methanol. (a) Comparison of catalytic activity. (b) Formation of products in scale-up methanol dehydrogenation over the Pd/Cd(Pd)S photocatalyst. (c) Comparison of $H_2$ formation over the Pd/Cd(Pd)S photocatalyst with presence of additives, 1mg of Pd/Cd(Pd)S was used. (d) Comparison of the AQYs of methanol dehydrogenation with other photocatalysts reported previously. The size of the circle is inversely proportional to the reaction temperature. (e) Correlation of wavelength-dependent AQYs of methanol dehydrogenation with UV-vis DRS spectrum of Pd/Cd(Pd)S. (f) Formation rate of HCHO on CdS and Pd/CdS photocatalysts in presence of electron sacrificial reagent (0.1 mmol). Standard reaction conditions: 1 mL of methanol, 2 mg of catalyst, 8 W blue LEDs (455 nm), Ar atmosphere, 1 h. Reaction conditions of measuring AQYs: 50 mL of methanol, 100 mg of Pd/Cd(Pd)S, 4 M of KOH, 20 mM of $Na_2S·9H_2O$, XX W LEDs (452

nm), Ar atmosphere, 0.5 h.

Methanol dehydrogenation over Cd(Pd)S produced $H_2$ with a low productivity of 9.2 mmol $g_{catal.}^{-1}$ and the amount of •$CH_2OH$ generated on Cd(Pd)S was higher than that on pristine CdS (Figure S11a). To distinguish the role of $Pd_1$ and Pd clusters, a Pd/CdS with a Pd loading amount of 0.1 at% (Pd/CdS(0.1)), equaling to that of Cd(Pd)S was prepared by photo-deposition. For the Pd/CdS(0.1) catalyst, although the productivity of $H_2$ (21.1 mmol $g_{catal.}^{-1}$) was 2.3 folds of Cd(Pd)S (Figure 5a), the formation rates of the total captured •$CH_2OH$ and •$OCH_3$ were close (1.16 and 1.26 mmol $g_{Cd(Pd)S}^{-1}$ $h^{-1}$, respectively). Moreover, methoxy radical (•$OCH_3$) was also generated with a 7% selectivity on the Cd(Pd)S, suggesting that the active sites of CdS and Cd(Pd)S are different as the O−H bond of methanol has a larger bond dissociation energy than that of C−H bond.[45] A possible explanation is that $Pd_1$ is at least one of the active sites for methanol oxidation. The activity of Cd(Pd)S maintained stability during the 8 h of reaction, indicating that $Pd_1$ is not transformed into Pd clusters as photogenerated holes are prone to be trapped by $Pd_1$, consistent with the conclusion inferred from DOS analysis. When $Pd_1$ and Pd clusters were present (Pd/Cd(Pd)S), $H_2$ productivity reached 332.7 mmol $g_{catal.}^{-1}$, equivalent to a TOF of 0.13 $s^{-1}$ based on Pd amount, which was 1.5 folds that of the Pd/CdS catalyst. To investigate if $Pd_1$ promotes the photocatalytic methanol dehydrogenation by reinforcing methanol adsorption, Cd(Pd)S and CdS were compared to assess if enhanced methanol adsorption contributes to the improved activity. As shown in Figure S12, the FT-IR spectra of methanol that adsorbed on CdS and Cd(Pd)S showed slightly weaker intensity, suggesting that $Pd_1$ on CdS has negligible promotional effects on the methanol dehydrogenation activity.

The carbonaceous products were quantified in a scale-up reaction (Figure 5b). At the initial stage of the reaction, HCHO was the major product and had a selectivity of > 99% (Figure S13). When prolonging the reaction time, the selectivity of CO increased slightly (~2%), accompanied by a decrease in $H_2$ evolution rate. This result was ascribed to the accumulation of HCHO: HCHO is more readily adsorbed on CdS than methanol,[46] leading to decrease of methanol dehydrogenation rate and promotion of CO generation. This result was also evidenced by Pd/CdS catalyst deactivation after generating almost the same amount of HCHO (Figure S14). Ethylene glycol (EG) was also produced but with very low productivity (0.1 mmol $g_{catal.}^{-1}$) and selectivity (< 0.1%) during 65 h of irradiation. The ratios of the amount of photogenerated electrons and holes consumed were between 0.89 and 1.08.

After reaction for 65 h, methanol conversion reached 4.0%. The spent Pd/Cd(Pd)S photocatalyst was studied means of XRD, TEM and ICP-AES. The XRD patterns of Pd/Cd(Pd)S showed similar diffraction peaks with that of the fresh catalyst (Figure S15a). The absence of Pd nanoparticles in the representative TEM images suggest that $Pd_1$ and Pd clusters did not grow up (Figures S15b and S15c). In addition, some of the CdS nanorods become smaller nanoparticles after the reaction as observed in the representative TEM image (Figure S15b), which may be caused by long-term stirring. However, CdS retained the hexagonal crystal phase after the reaction (Figure S15c). The ICP-AES results indicated a Pd content of 0.8 at% for the spent Pd/Cd(Pd)S catalyst after the reaction. Based on the above results, the main reason for the deactivation of the Pd/Cd(Pd)S catalyst is the leaching of Pd clusters. After supplementing Pd clusters by photo-deposition method. The activity of the Pd/Cd(Pd)S catalyst almost recovered and was even slightly higher than the fresh catalyst (Figure S15d).

The hydrogen kinetic isotope effect (KIE) experiments indicated that methanol C−H bond cleavage was involved in the rate-determining step. When $CH_3OH$ and $CD_3OH$ were used as substrates, the KIE values based on the productivity of HCHO were 2.9 and 3.1, respectively, for Pd/Cd(Pd)S and Pd/CdS (Figure S16). In this regard, 2 M of KOH was added as a proton acceptor to promote proton-coupled electron transfer (Figure 5c).[47] Furthermore, considering that formation of S vacancies can suppress $H_2$ evolution reaction,[48] electron paramagnetic resonance (EPR) spectroscopy investigation was performed. Observing a signal with a g value of 2.003 suggests formation of S vacancies in the spent Pd/Cd(Pd)S catalyst when KOH was used (Figure S17a). In order to compensate for the negative effects of S vacancy formation (Figure S17b), 10 mM of $Na_2S$ was added to the reaction solution (Figure 5c). EPR spectrum of the spent Pd/Cd(Pd)S recycled from the suspension containing both KOH and $Na_2S$ fully confirmed suppression of the S vacancy formation since the signal with a g value of 2.003 disappeared (Figure S17a). The addition of KOH and $Na_2S$ improved the productivity of $H_2$ to 4.1 times for the same reaction time. Comparison experiments suggest that the $H_2$ was derived from methanol dehydrogenation since the amount of produced $H_2$ was 5.1 mmol $g_{catal.}^{-1}$ without methanol for a reaction time of 1 h (Figure 5c), much lower than that (208.4 mmol $g_{catal.}^{-1}$) if methanol and $Na_2S$ were present. Methanol was dehydrogenated with a TOF of 0.64 $s^{-1}$, and was improved to 1.14 $s^{-1}$ at an input LEDs power of 16 W (Figure S18). The AQYs for $H_2$ and HCHO production were both 87 ± 1% when KOH and $Na_2S$ were present (Figure 5d and Table S5), confirming that $Na_2S$ was only subtly oxidized. Notably, KOH can catalyze the Cannizzaro reaction of HCHO with coproduction of formic acid and methanol. However, because the reaction temperature was low (298 K), the AQY of formic acid was lower than 0.1%. Based on the optimized reaction conditions, the measured AQYs at 365, 425, 452 and 525 nm were 55 ± 3, 80 ± 3, 87 ± 1 and 26 ± 1%, respectively, almost consistent with the absorption spectrum of the Pd/Cd(Pd)S catalyst (Figure 5e). The lower AQY at 365 nm may be due to the formation of S vacancies even $Na_2S$ was used to remedy them.[20]

To explain why the oxidation half-reaction was accelerated by loading Pd SAs on CdS, we selected proper electron scavenger to rapidly remove electrons, thus the whole reaction is limited by the reaction of holes. A large number of references prove that $CCl_4$ (*Appl. Catal.*, B 2020, 271, 11894; *J. Hazard. Mater.*, 2019, 367, 277) and $K_2S_2O_8$ (*ACS Catal.*, 2022, 12, 8, 4481–4490; *J. Am. Chem. Soc.* 2022, 144, 8, 3386–3397; *ACS Energy Lett.*, 2019, 4, 1, 203–208) can be quickly scavenge electrons under conditions similar with the photocatalytic methanol dehydrogenation in this manuscript. Although 2-phenoxy-1-phenylethan-1-one (PP-one) is not commonly used as an electron-capturing reagent, research on the hydrogenolysis of PP-one and the dehydrogenation-hydrogenolysis of PP-ol has shown that electrons on CdS can reduce the C–O bond of PP-one (*ACS Catal.* 2017, 7, 7, 4571–4580). With the addition of $CCl_4$, $K_2S_2O_8$, and PP-one, the formation rate of HCHO increased by 2.8, 1.4, and 1.7 folds (Figure 5f), respectively. Because $Pd_1$ promoted hole transfer to CdS surface as revealed by photocurrent density (Figure 4d), the increased HCHO formation rate was ascribed to catalytic methanol oxidation on the $Pd_1$ site. The overall experimental results supported a Janus Pd that promotes methanol oxidation and $H_2$ evolution. Furthermore, we found that other metals like Ag and Cu can exchange with $Cd^{2+}$ and also promote methanol oxidation, similar to that of $Pd_1$ (Figure S19), indicating that the synergistic effect of single sites and clusters can be extended to other elements.

**Conclusions**

In summary, we achieved high-quantum-efficient methanol dehydrogenation capitalizing on the complexity of atomically dispersed Pd and clustered Pd. The $Pd_1$ and Pd clusters were loaded on CdS by sequential ion exchange and photo-deposition methods. The structure of the two kinds of Pd was evidenced by XAFS spectroscopy and theoretical methods. $Pd_1$ in substitution of $Cd^{2+}$ on CdS (100) crystal plane forms defect energy levels that trap photogenerated holes for methanol oxidation. The $Pd_1$ also assists the dispersion of Pd clusters prepared by photo-deposition in methanol solution. The

Pd$_1$ and Pd clusters promote methanol oxidation and H$_2$ evolution, respectively, originating a synergistic catalysis for photocatalytic methanol dehydrogenation. Therefore, methanol was dehydrogenated with a highest TOF of 1.14 s$^{-1}$ over the Pd/Cd(Pd)S photocatalyst, affording H$_2$ and HCHO with similar a AQY of 87 ± 1% at 452 nm. The synergistic Pd$_1$ and Pd clusters promise highly efficient coproduction of H$_2$ and HCHO capitalizing on the synergistic catalysis of supported metal species.

**Methods**

**Experimental Methods.**

**Preparation of catalysts.** CdS nanorods were synthesized through a hydrothermal method according to the reported procedures.[55] Typically, Cd(NO$_3$)$_2$·4H$_2$O (2 mmol) and thiourea (8 mmol) were dissolved in 20 mL of ethylenediamine anhydrous in a 50 mL breaker and magnetically stirred until the solids dissolved completely at room temperature. Then, the mixture was transferred to a 50-mL stainless Teflon-lined autoclave, tightly sealed and placed in a 160 °C oven for 20 h. The system was then naturally cooled to room temperature. After being washed with deionized water (3 × 25 mL) and absolute ethanol (3 × 25 mL), a yellow solid was obtained after being dried under vacuum at 60 °C for 12 h.

Cd(Pd)S catalysts were prepared simply by the cation exchange method as follows: a certain amount of H$_2$PdCl$_4$ aqueous solution (11.1 mM) was added to the suspension of CdS (50 mg) dispersed in deionized water (50 mL). Then, the mixture was stirred for 24 h in the dark at room temperature and atmosphere. The resulting yellow slurry was slightly dark compared with pure CdS nanorods. After being washed with deionized water (3 × 25 mL) and absolute ethanol (3 × 25 mL), Cd(Pd)S catalysts were obtained after being dried under vacuum at 60 °C for 12 h. The preparation methods of Cd(Ag)S and Cd(Cu)S are similar, except that the precursor solution is replaced with AgNO$_3$ and Cu(NO$_3$)$_2$ aqueous solutions.

Pd/Cd(Pd)S were prepared by photo deposition method. Typically, 100 mg of Cd(Pd)S and a certain amount of H$_2$PdCl$_4$ aqueous solution were added into 50 mL of CH$_3$OH in a quartz reactor. After ultrasonication to remove O$_2$, the system was replaced with Ar for 5 min. Then the reactor was sealed and placed under light illumination for 1 h under stirring. Then, the precipitate was separated centrifugally. After being washed with deionized water (3 × 25 mL) and absolute ethanol (3 × 25 mL), Pd/Cd(Pd)S heterostructures were obtained after being dried under vacuum at 60 °C for 12 h. The preparation method of Pd/CdS was similar to the above process, except that the support added during photo-deposition was CdS instead of Cd(Pd)S.

**Photocatalytic experiments.** Photocatalytic dehydrogenation of methanol was carried out in homemade LED photoreactors with a total power of 8 W (455 nm). Typically, photocatalysts and 1 mL of methanol were added into a quartz tube with a volume of 4 mL. The atmosphere in the quartz tube was then replaced by Ar and tightly sealed. The quartz tube was installed in the photoreactor and allowed for light illumination for a desired time under stirring. The scale-up reaction was carried out in a top-illuminated reactor with blue LEDs (input power of 87 W, 452 ± 10 nm). 50 mg of Pd/Cd(Pd)S was dispersed in 50 mL of methanol, 100 μL of gas and 500 μL of liquid were sampled per analysis. The AQYs of the photocatalytic dehydrogenation of methanol were measured over Pd/Cd(Pd)S and Pd/CdS with blue LEDs (input power of 27.2 W, 452 ± 10 nm). The experiment was carried out in a homemade photoreactor with a 6 cm diameter by top irradiation. 100 mg of the photocatalyst was dispersed in 50 mL of methanol. The temperature was kept at ~25 °C by cycle cooling water. The number of photons reaching the top of the reaction solution was measured with a calibrated Si photodiode (LS-100, EKO Instruments). The AQYs ($\eta$) for the hydrogen production were calculated using the following equation:

$$\eta = \frac{2n_{H2}(\text{mol}) \times N_A(\text{mol}^{-1})}{I(\text{cm}^{-2}\text{s}^{-1}) \times t(\text{s}) \times S(\text{cm}^2)}$$

where $n_{H2}$ is the production of hydrogen, and $N_A$, $I$, $t$ and $S$ represent Avogadro's constant, light intensity, reaction time and irradiation area, respectively.

**Photoelectrochemical measurements**

Transient photocurrent density curves were measured on a CS2350H instrument in the three-electrode system. Ag/AgCl and Pt mesh electrodes were used as the reference and counter electrodes. The working electrodes were prepared by coating as-prepared samples on indium-doped tin oxide (ITO) glass electrodes. The photocurrent was performed at a potential of 0 V versus Ag/AgCl in a 0.1 M solution of tetrabutylammonium hexafluorophosphate in methanol or acetonitrile. LED light with 455 nm was applied as the light source.

**DFT calculations**

Spin-polarized DFT calculations were performed using the Vienna Ab Initio Package (VASP), where the projected augmented wave (PAW) pseudopotentials were applied to describe the ionic cores. The valence electrons were explicitly considered using a plane-wave basis set with energy cutoffs of 450 eV throughout the study. The generalized gradient approximation (GGA) using the PBE functional was applied, and Grimme's DFT-D3 methodology was used to describe the dispersion interactions among all the atoms.

The primitive cell of CdS was first optimized where the Brillouin zone was sampled using 9 × 9 × 9 Γ-centered Monkhorst−Pack grid meshes. The slabs with (100) and (001) facet surfaces were constructed using the optimized primitive cell with further structure optimization due to Pd doping. The slabs were separated by a vacuum layer of 15 Å to mitigate the interaction between the slab and its periodic replicas. In all calculations related to the slabs, the bottom two layers were fixed to their bulk configurations, while the top layers and the adsorbates were free to relax in all directions. Brillouin zone integration was performed using Gaussian smearing of 0.05 eV. All self-consistent field (SCF) calculations were converged to $1 \times 10^{-5}$ eV. All atomic coordinates of the adsorbates and the metal atoms in the top two layers were optimized to a force of less than 0.02 eV/Å on each atom. The density of state (DOS) of the optimized geometries was calculated using the HSE06 exchange-correlation function implemented.

**XANES/EXAFS Spectroscopy**

XAFS experiments at the Pd K edge have been performed at SAMBA beamline of synchrotron SOLEIL.[56] Two Pd mirrors have been used to reject harmonics, the monochromator was equipped with Si (220) crystals, and measurements have been performed in fluorescence mode with a 35 pixels HPGe detector (Mirion/Canberra) in conjunction with DxMap DSP (XIA). Measurements of reference materials have been performed in transmission. In order to refine EXAFS, the XAFS spectra have been analyzed using the DEMETER suite (Athena to extract signal, Artemis to refine the fit). The $S_0^2$ has been deduced from a fitting on a palladium foil.

*Ex-situ* samples were analysed after deposition of the powders on PVDF membrane. Spectra for *in situ* irradiated samples were acquired employing conditions very similar to that used for the photocatalytic tests (see below). Briefly, 10 mg of the photocatalyst was suspended in 1.0 mL of $CH_3OH$ within a quartz vial and closed with a PTFE/silicone. After removal of air by an Ar stream, the vial was maintained under stirring and the photocatalyst was irradiated with a 450 nm diode emission laser, with an irradiation power of 100 mW cm$^{-2}$ over a spot with a diameter of 0.5 cm$^2$. Irradiation was performed for 4 h. After that time, the vial was transferred inside a glove box to avoid exposure to air, filtered the sample on PVDF membrane and sealed for analysis within two Kapton tapes.

## Associated Content

Supporting Information: including detailed experimental procedures, XRD patterns, TEM images, UV-Vis DRS, PL spectra, TPC curves, GC and MS results.

## Author Information

Corresponding Authors

Nengchao Luo − State Key Laboratory of Catalysis, Dalian National Laboratory for Clean Energy, Dalian Institute of Chemical Physics, Chinese Academy of Sciences, Dalian, 116023, China; orcid.org/0000-0002-6137-292X;

Email: ncluo@dicp.ac.cn

Paolo Fornasiero − Department of Chemical and Pharmaceutical Sciences, CNR-ICCOM Trieste and INSTM Trieste Research Units, University of Trieste, via L. Giorgieri 1, 34127 Trieste, Italy; orcid.org/0000-0003-1082-9157

Email, pfornasiero@units.it

Feng Wang − State Key Laboratory of Catalysis, Dalian National Laboratory for Clean Energy, Dalian Institute of Chemical Physics, Chinese Academy of Sciences, Dalian, 116023, China; orcid.org/0000-0002-9167-8743;

Email: wangfeng@dicp.ac.cn

Authors

Zhuyan Gao − State Key Laboratory of Catalysis, Dalian National Laboratory for Clean Energy, Dalian Institute of Chemical Physics, Chinese Academy of Sciences, Dalian, 116023, China; orcid.org/0000-0001-5965-3901;

Tiziano Montini − Department of Chemical and Pharmaceutical Sciences, CNR-ICCOM Trieste and INSTM Trieste Research Units, University of Trieste, via L. Giorgieri 1, 34127 Trieste, Italy; orcid.org/0000-0001-9515-566X;

Junju Mu − State Key Laboratory of Catalysis, Dalian National Laboratory for Clean Energy, Dalian Institute of Chemical Physics, Chinese Academy of Sciences, Dalian, 116023, China; orcid.org/0000-0002-7382-2182;

Emiliano Fonda − Synchrotron SOLEIL, L'Orme des Merisiers, BP48 Saint Aubin, 91192 Gif-sur-Yvette, France; orcid.org/0000-0001-6584-4587.

**Notes**

The authors declare no competing interests.

**Acknowledgments**

This work is supported by the National Key R&D Program of China (2022YFA1504904), the National Natural Science Foundation of China (22025206, 21991090, 22202199, 22172157), the Dalian Innovation Support Plan for High Level Talents (2022RG13), the Youth Innovation Promotion Association（YIPA）of the Chinese Academy of Sciences (2023192), the Fundamental Research Funds for the Central Universities (20720220008), the Foreign Expert Project (G2022008003), the Postdoctoral Fellowship Program of CPSF (GZB20240724). We also thank the instrumental support of the Liaoning Key Laboratory of Biomass Conversion for Energy and Material. P.F. and T.M. acknowledge European Union (projects HORIZON-WIDERA-2021-ACCESS-03-01–Grant No. 101079384 and HORIZON-EIC-2023-PATHFINDEROPEN-01-01: Grant No. 101130717) for financial support.

# Supporting Information for:

# Photocatalytic methanol dehydrogenation promoted synergistically by atomically dispersed Pd and clustered Pd

Zhuyan Gao,[1,2,5] Tiziano Montini,[3,5] Junju Mu,[1] Nengchao Luo,[1,*] Emiliano Fonda,[4] Paolo Fornasiero[3,*] and Feng Wang[1,2,*]

[1] State Key Laboratory of Catalysis, Dalian National Laboratory for Clean Energy, Dalian Institute of Chemical Physics, Chinese Academy of Sciences, Dalian, 116023, China.

[2] University of Chinese Academy of Sciences, Beijing, 100049, China.

[3] Department of Chemical and Pharmaceutical Sciences, Center for Energy, Environment and Transport Giacomo Ciamiciam, INSTM Trieste Research Unit and ICCOM-CNR Trieste Research Unit, University of Trieste, Via Licio Giorgieri 1, 34127 Trieste, Italy

[4] Synchrotron SOLEIL, L'Orme des Merisiers, Saint Aubin BP48, 91192 Gif sur Yvette CEDEX, France

[5] These authors contributed equally: Zhuyan Gao, Tiziano Montini.

* Corresponding Author.

*N. L.: e-mail, ncluo@dicp.ac.cn

*P. F.: e-mail, pfornasiero@units.it

*F. W.: e-mail, wangfeng@dicp.ac.cn

# Table of Content

**Supporting Methods**

**Materials and reagents**

All chemicals were commercially available and used without further purification. $Cd(NO_3)_2 \cdot 4H_2O$ (AR) and ethylenediamine anhydrous (AR) were purchased from Kermel Reagent Co., Ltd. Thiourea (99%) was purchased from Aladdin-Reagent. Metal salts were purchased from Shenyang Research Institute of Nonferrous Metal Co. Ltd., methanol was purchased from Sinopharm Chemical Reagent Co., Ltd.

**Quantitative analysis of products**

$H_2$ and CO were quantified with He as the internal standard. After the reaction, the gas products were qualitatively analysed by Techcomp GC 7900 gas chromatograph (TCD detector, TDX-01 column) with Ar as the carrier gas. The produced $H_2$ can be calculated from the following equation:

$$n(H_2) = \frac{0.5529 \times A(H_2) \times V(He)/mL}{A(He)} \times \frac{101.3}{8.314 \times 298} mmol = 0.0226 \times \frac{A(H_2)}{A(He)} \times \frac{V(He)}{mL} \ mmol \qquad (1)$$

$$n(CO) = \frac{5.3562 \times A(CO) \times V(He)/mL}{A(He)} \times \frac{101.3}{8.314 \times 298} mmol = 0.2190 \times \frac{A(CO)}{A(He)} \times \frac{V(He)}{mL} \ mmol \qquad (2)$$

where $n(H_2)$ is the produced $H_2$ in the reaction system, mmol; $A(H_2)$ and $A(He)$ are the peak area of $H_2$ and He, respectively. $V(He)$ is the injected volume of He. The produced $H_2$ then expresses as $n(H_2)/m_{catalyst}$, here, $m_{catalyst}$ represents the mass of the catalyst.

HCHO was quantified by a derivatization method with *p*-chloroanisole as the internal standard. After reaction, the residual reaction mixture was filtrated through a 0.22 μm Nylon syringe filter to remove the catalyst. 50 μL of reaction solution and 50 μL of an internal ethanol solution containing *p*-chloroanisole was mixed, which was diluted to 2 mL with MeCN. Then 50 μL of the diluted reaction mixture, 0.70 mL of 2,4-dinitrophenylhydrazine (DNPH)/MeCN/$H_2O$ solution (prepared by dissolving 0.10 g of DNPH and 4.0 mL of concentrated hydrochloric acid into 50 mL of $CH_3CN$ followed by diluting the solution to 100 mL with deionized water) and 0.25 mL of deionized water were mixed and heated at 50 °C for 1 h to totally convert HCHO into HCHO-DNPH. The reaction equation is as follows:

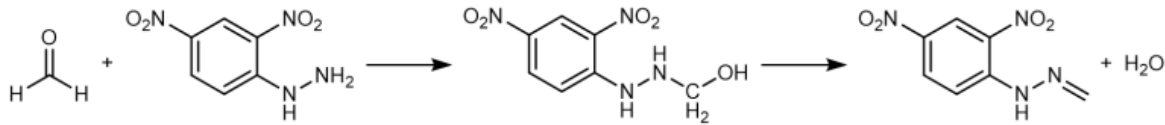

The mixture was then analysed by high pressure liquid chromatography (HPLC, waters XSelect HSS-PFP column, maintained at 30 °C, UV-detector at 232 nm, mobile phase: 45% $CH_3CN$ balanced by $H_2O$, 1.0 mL min$^{-1}$).

HCOOH was quantified with propanol as the internal standard and analyzed by HPLC (Agilent 1260 Infinity, equipped with Hi-Plex H column (300 mm × 7.7 mm) and RID-6A refractive index detector) with diluted 0.5 mM $H_2SO_4$ solution as the mobile phase.[1] A UV detector recorded at 210 nm was used to analyze HCOOH.

Other liquid products were quantitatively analysed with 1,3-propanediol as the internal standard by gas chromatography equipped with a flame ionization detector (GC-FID, Agilent 7890A, column: HP-5, 30 m × 530 μm × 1.5 μm). The productivity of product, selectivity, and n(h$^+$)/n(e$^-$) were calculated according to the following equations:

$$\text{Productivitiy of product} = \frac{n_{\text{product}}}{m_{\text{catalyst}}} \times 100\% \tag{3}$$

$$S_{\text{HCHO}} = \frac{n_{\text{HCHO}}}{n_{\text{HCHO}} + n_{\text{CO}} + 2n_{\text{EG}}} \times 100\% \tag{4}$$

$$S_{\text{CO}} = \frac{n_{\text{CO}}}{n_{\text{HCHO}} + n_{\text{CO}} + 2n_{\text{EG}}} \times 100\% \tag{5}$$

$$S_{\text{EG}} = \frac{2n_{\text{EG}}}{n_{\text{HCHO}} + n_{\text{CO}} + 2n_{\text{EG}}} \times 100\% \tag{6}$$

$$n_{\text{h+}}/n_{\text{e-}} = \frac{2n_{\text{HCHO}} + 4n_{\text{CO}} + 2n_{\text{EG}}}{2n_{\text{H2}}} \tag{7}$$

$n_{\text{product}}$ represents the molar amount of corresponding product, $m_{\text{catalyst}}$ represents the mass of the used catalyst. $S$ represent the selectivity of corresponding product. $n_{\text{h+}}$ and $n_{\text{e-}}$ represent the number of moles of holes and electrons participating in the photocatalytic reaction, respectively.

**AQY measurements.** The AQY of photocatalytic methanol dehydrogenation were measured over Pd/Cd(Pd)S and Pd/CdS with blue LEDs (452 nm ± 10 nm, 27.4 W), respectively. The number of photons reaching the top of the reaction solution was measured with a calibrated Si photodiode (LS-100, EKO Instruments). The AQYs ($\eta$) for hydrogen production were calculated using the following equation:

$$\eta = \frac{2n_{H2}\ (\text{mol}) \times N_A\ (\text{mol}^{-1})}{I\ (\text{cm}^{-2}\text{s}^{-1}) \times t\ (\text{s}) \times A\ (\text{cm}^2)} \times 100\%$$

where $n_{H2}$ is the formed hydrogen, and $N_A$, $I$, $t$, and $A$ represent Avogadro's constant, number of incident photons, reaction time and irradiation area, respectively.

**General characterization**

**Inductively coupled plasma optical emission spectroscopy (ICP-OES).** ICP-OES were conducted with ICP-OES 7300DV Spectrometer (PerkinElmer). 10 mg of CdS was dispersed in 5 mL of water and methanol, respectively. After stirring 3 h in dark, the ion concentration in the filtrate was measured.

**X-ray diffraction (XRD).** Powder X-ray diffraction patterns (XRD) were conducted with a PANalytical X-Pert PRO diffractometer, using Cu Kα radiation at 40 kV and 20 mA. The data were recorded over a 2θ range of 20-80°.

**Transmission electron microscopy (TEM).** Samples for TEM were prepared by dispersing Au/CdS in ethanol and ultrasonicating for 20 min. The suspension was loaded onto a Cu TEM grid and dried. TEM images were obtained using a JEM-F200 field emission transmission electron microscope at an accelerating voltage of 200 kV.

**UV-vis diffuse reflectance spectra.** Catalysts for UV–vis diffuse reflectance spectra were recorded on a SHIMADZU UV-2600 spectrophotometer in a wavelength range of 300-800 nm with BaSO$_4$ as the background.

**Diffuse reflectance infrared Fourier transform spectroscopy (DRIFTS).** DRIFTS was measured on a Thermo Scientific Nicolet iS50 Adv FTIR Spectrometer equipped with a diffuse reflectance accessory and a liquid nitrogen-cooled MCT detector. In a typical measurement procedure, the sample cell containing the powder sample was connected to gas lines that can switch gases between 5% CO in Ar and Ar. The sample was primarily treated by Ar, and after several minutes, 5% CO in Ar was introduced and kept for 10 min, followed by switching to Ar and collecting signals. When measuring the methanol adsorption IR spectrum, methanol-containing Ar was introduced by bubbling.

**X-ray photoelectron spectrum (XPS).** XPS was recorded on a Thermofisher Escalab 250Xi+ instrument, using monochromated Al Kα radiation as the X-ray source. The binding energies were calibrated according to the position of C 1s adventitious carbon.

**Photoluminescence spectra (PL).** PL spectra were measured on a Hitachi F-7000 Fluorescence Spectrophotometer using a Xenon lamp as the excitation source at room temperature.

**Electron paramagnetic resonance (EPR).** EPR spectra were measured on a Bruker spectrometer in the X-band at 77 K with a field modulation of 100 kHz. Samples for measurements were obtained by filtrating the solid from the reaction mixture in a glove box (argon atmosphere). The samples were then loaded into paramagnetic tubes and sealed in a glove box to protect the samples.

**Supporting Figures**

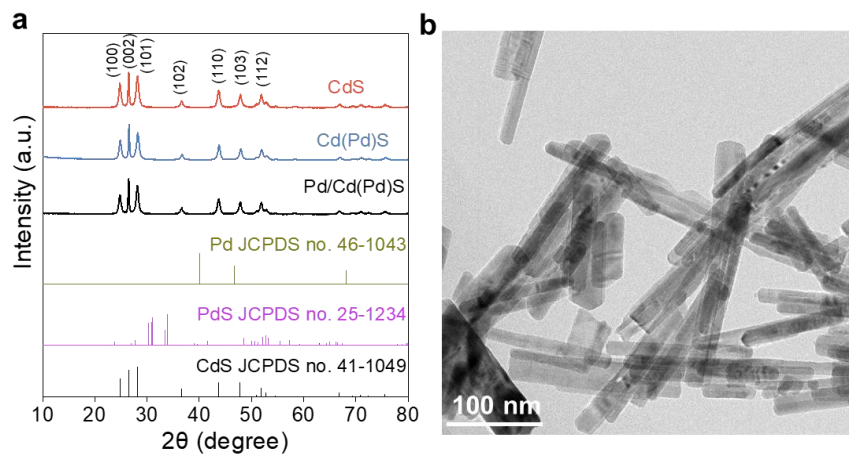

**Figure S1.** Characterization of prepared catalysts. (**a**) XRD patterns of prepared catalysts. (**b**) TEM image of Pd/Cd(Pd)S.

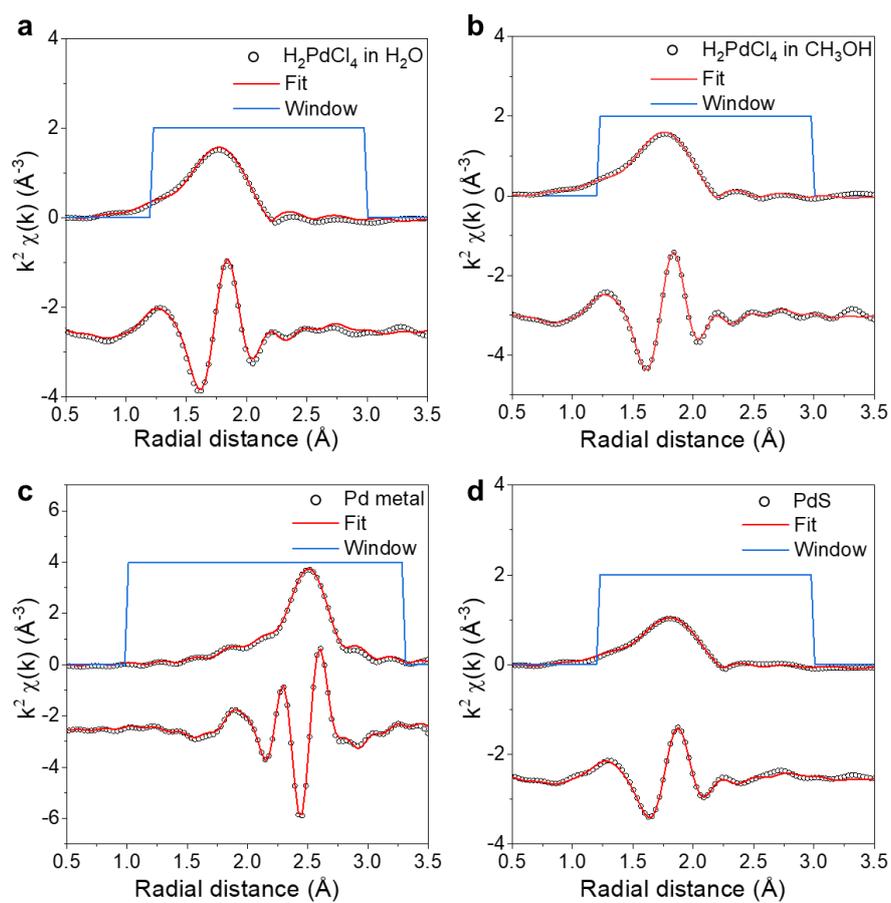

**Figure S2.** Fits of FT of k$^2$χ(*k*) of the EXAFS signals of H$_2$PdCl$_4$ in aqueous (**a**) and methanolic solution (**b**). Fits of FT of k$^2$χ(*k*) of the EXAFS signals of Pd metal (**c**) and PdS (**d**).

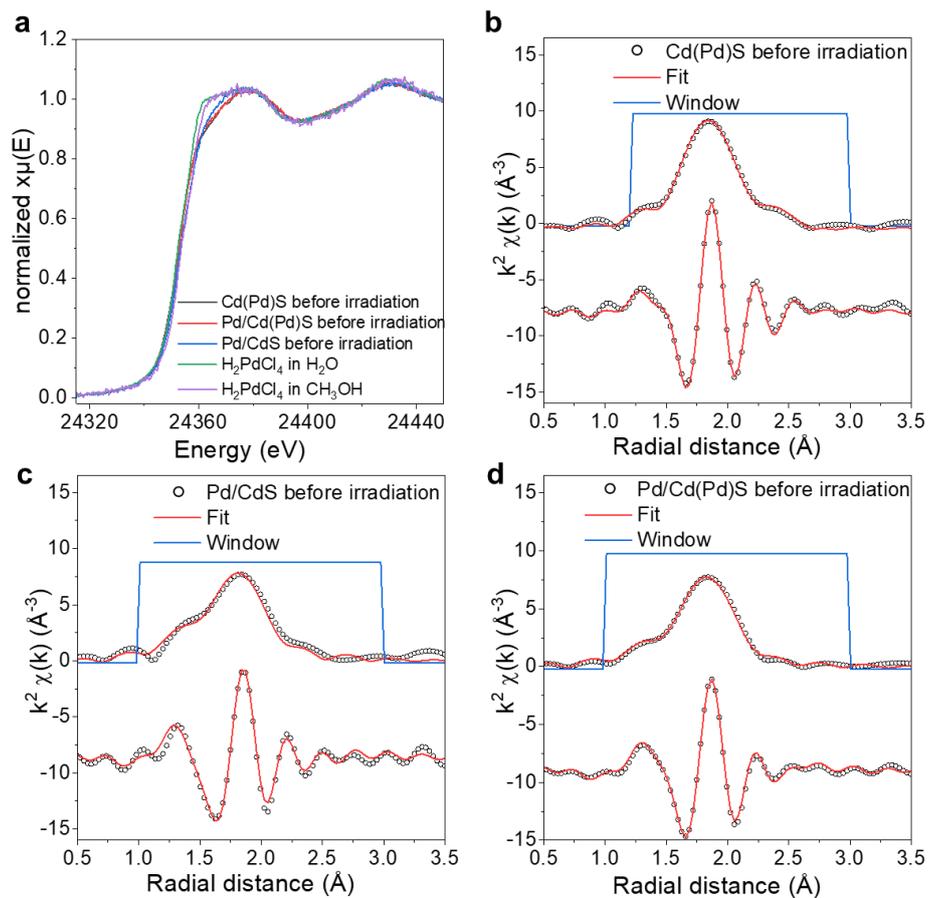

**Figure S3.** (**a**) Pd K-edge XANES spectra of samples before irradiation. Fits of FT of k² χ(k) of the EXAFS signals of Cd(Pd)S (**b**), Pd/CdS (**c**), and Pd/Cd(Pd)S (**d**) before irradiation, respectively.

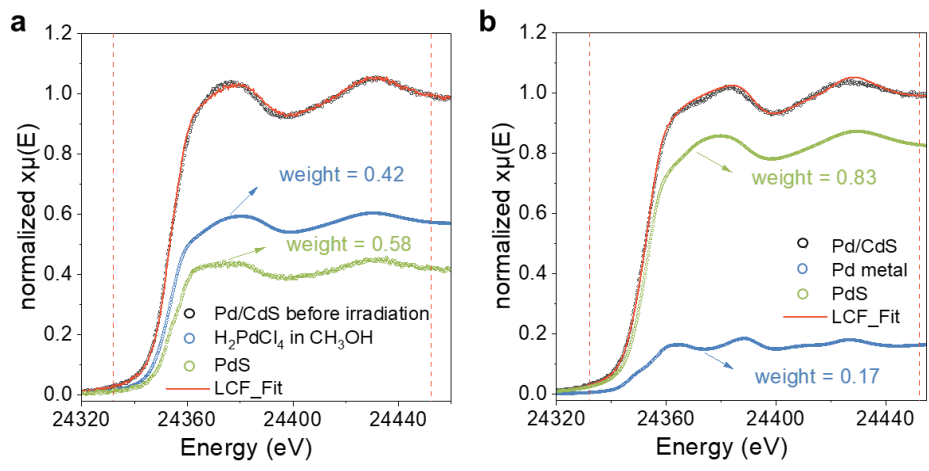

**Figure S4.** Linear combination fitting (LCF) of Pd/CdS XANES before (**a**) and after irradiation (**b**).

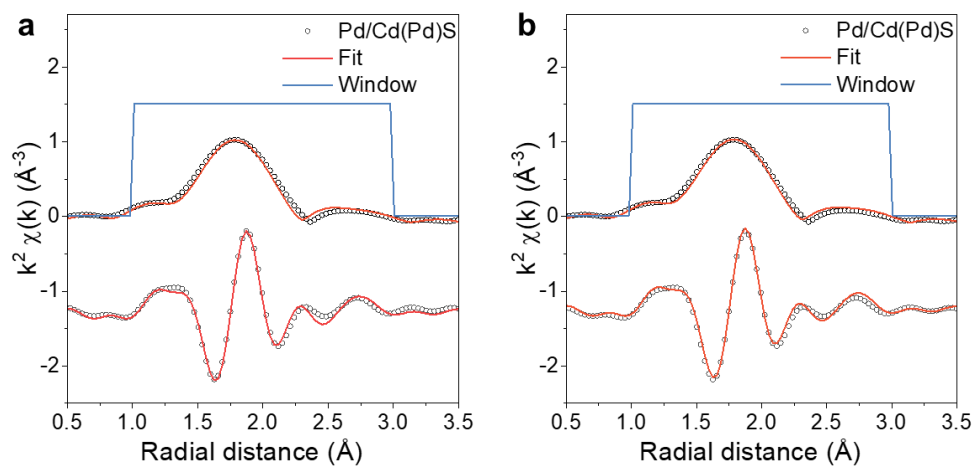

**Figure S5.** Fitting of the FT $\chi(k)$ of the EXAFS signals of Pd/Cd(Pd)S after irradiation in $CH_3OH$ using different models.

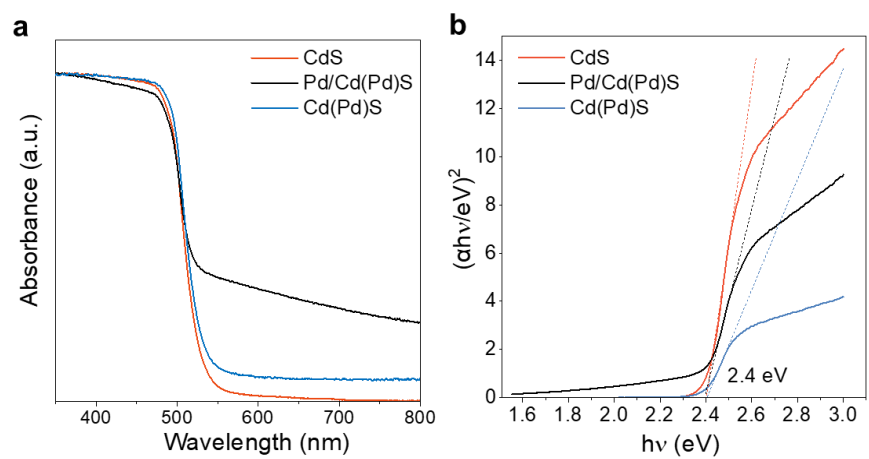

**Figure S6.** (**a**) UV-vis diffuse reflectance spectra of samples. (**b**) Plot of transformed Kubelka–Munk function.

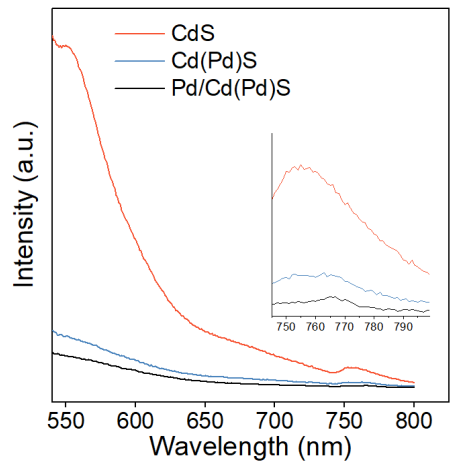

**Figure S7.** Steady-state photoluminescence spectra of samples.

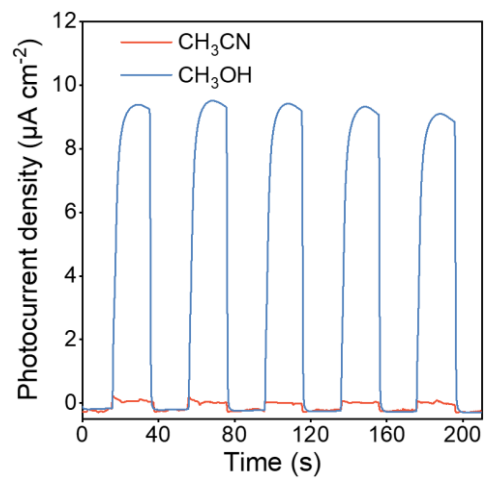

**Figure S8.** Transient photocurrent density curves of Pd/Cd(Pd)S in $CH_3CN$ and $CH_3OH$.

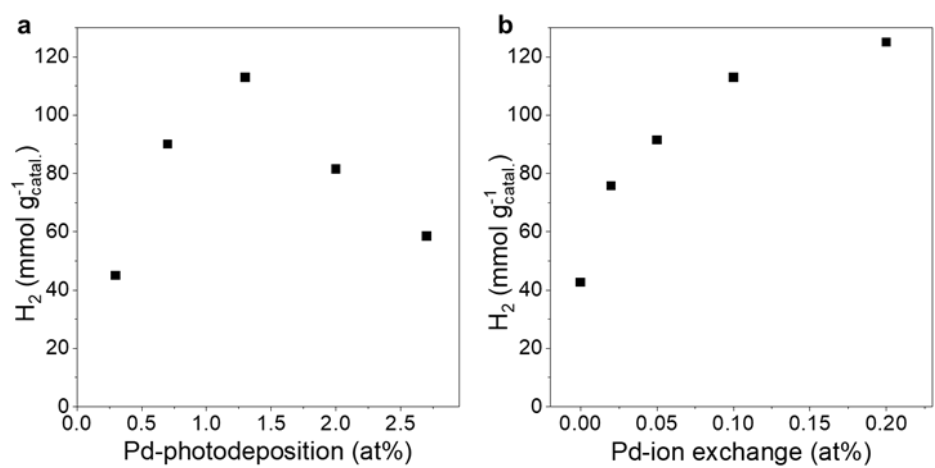

**Figure S9.** Optimization of Pd loadings. Influence of loadings of photo-deposited (**a**) and ion-exchanged Pd (**b**) on $H_2$ evolution from methanol dehydrogenation. Standard reaction conditions: 1 mL of methanol, 2 mg of CdS or Cd(Pd)S, 8 W blue LEDs (455 nm), Ar atmosphere, 2 h.

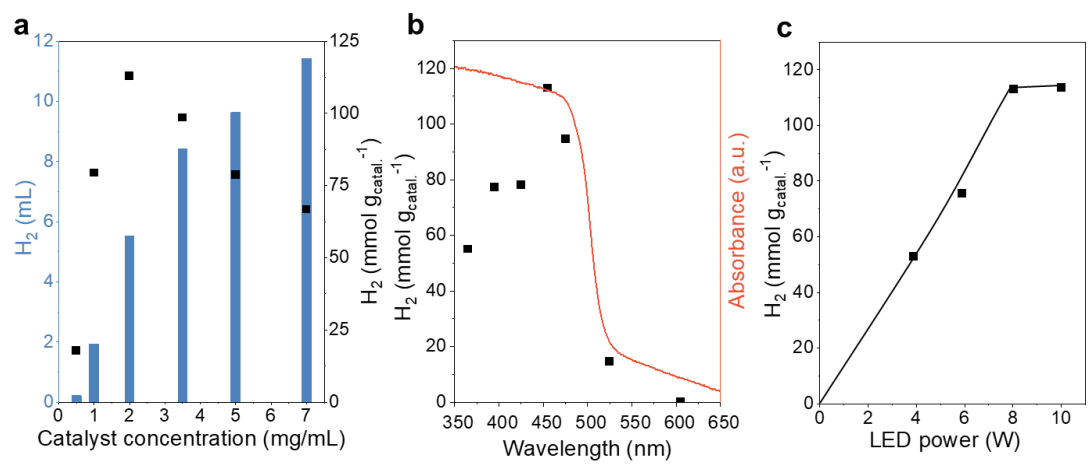

**Figure S10.** Optimization experiments of catalyst concentration (**a**), light wavelength (**b**) and intensity (**c**). Standard reaction conditions: 1 mL of methanol, 2 mg of Pd/Cd(Pd)S, 8 W blue LEDs (455 nm), Ar atmosphere, 2 h.

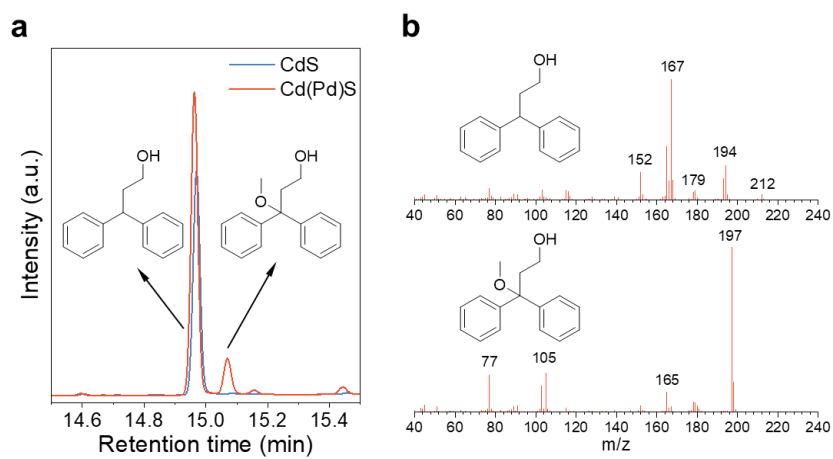

**Figure S11. (a)** GC spectrum of radical capturing experiment using Pd/Cd(Pd)S photocatalyst. **(b)** Mass spectrograms of products in radical capturing experiments with retention times of 14.97 and 15.06 min, respectively.

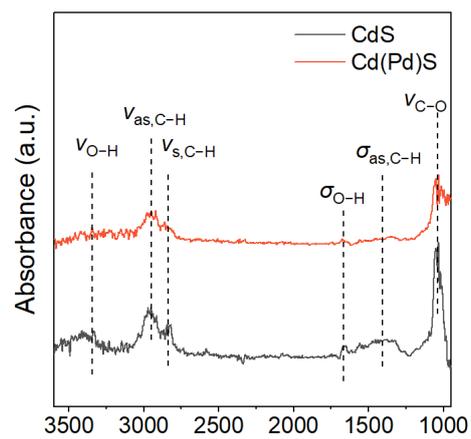

**Figure S12.** FT-IR spectra of methanol adsorbed on CdS and Cd(Pd)S, respectively.

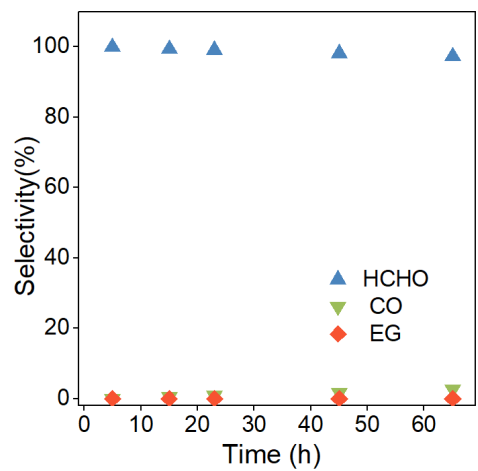

**Figure S13.** Selectivity of carbon-containing products in scale-up experiment over the Pd/Cd(Pd)S photocatalyst. Standard reaction conditions: 50 mL of methanol, 50 mg of Pd/Cd(Pd)S, 87 W blue LEDs (452 nm), Ar atmosphere.

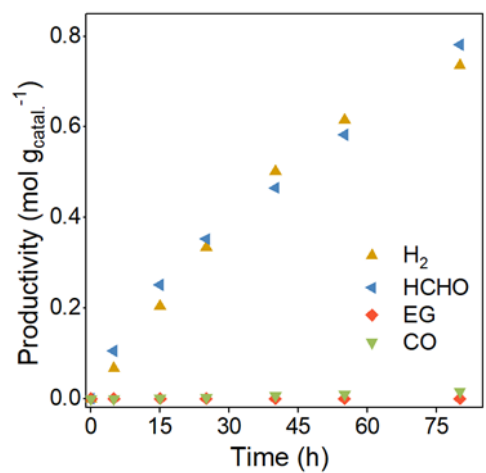

**Figure S14.** Formation of products in a scale-up photocatalytic methanol dehydrogenation on Pd/CdS catalyst. Standard reaction conditions: 50 mL of methanol, 50 mg of catalysts, 87 W blue LEDs (452 nm), Ar atmosphere.

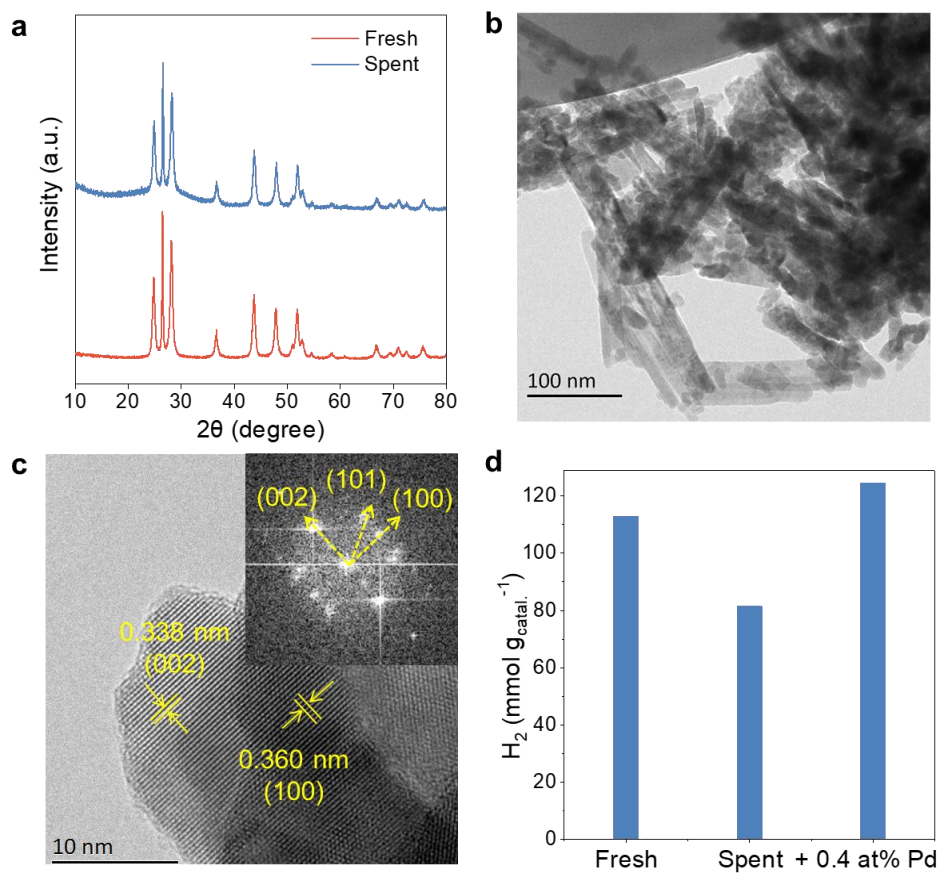

**Figure S15.** Characterization of the spent Pd/Cd(Pd)S catalyst. (**a**) XRD patterns of fresh catalysts and spent catalysts. (**b**, **c**) Representative TEM images of used catalysts (**d**) Regeneration of used catalysts. Standard reaction conditions: 1 mL of methanol, 2 mg of catalysts, 8 W blue LEDs (455 nm), Ar atmosphere, 2 h.

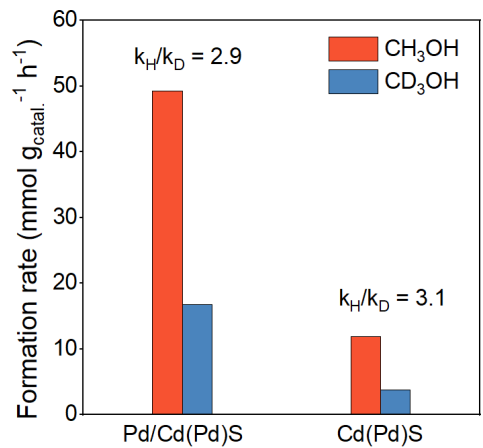

**Figure S16.** Hydrogen KIE experiments. Reaction conditions: 1 mL of CH$_3$OH or CD$_3$OH, 5 mg of photocatalysts, 8 W blue LEDs (455 nm), Ar, 3h.

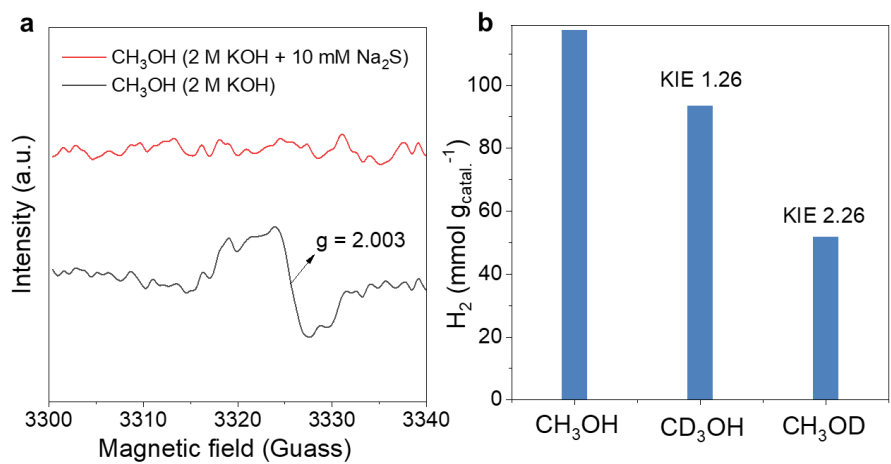

**Figure S17.** Optimization of additives to promote photocatalytic methanol dehydrogenation over the Pd/Cd(Pd)S photocatalyst. (**a**) EPR spectra of Pd/Cd(Pd)S with additives after reaction. (**b**) KIE experiments to find the rate-limiting step of methanol dehydrogenation in the presence of KOH. Standard reaction conditions: 1 mL of methanol, 2 mg of Pd/Cd(Pd)S, 8 W blue LEDs (455 nm), Ar atmosphere, 1 h.

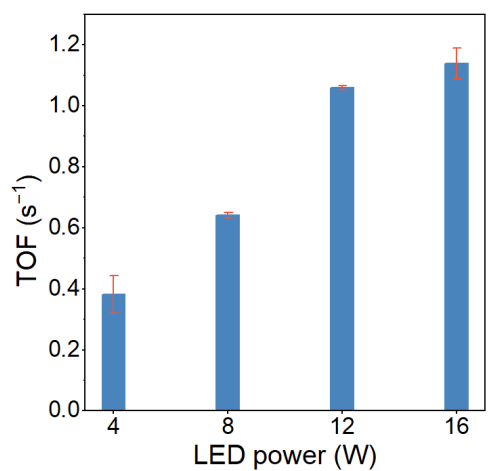

**Figure S18.** Optimization experiments of light intensity. Standard reaction conditions: 1 mg of Pd/Cd(Pd)S, 2 M of KOH, 10 mM of Na$_2$S·9H$_2$O, 455 nm LEDs, Ar atmosphere, 1 h.

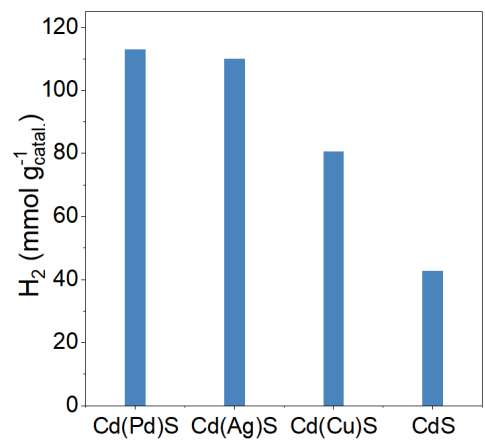

**Figure S19.** Comparison of the activity of Pd/Cd(M)S and Pd/CdS for photocatalytic methanol dehydrogenation. Standard reaction conditions: 1 mL of methanol, 2 mg of catalysts, 8 W blue LEDs (455 nm), Ar atmosphere, 2 h. M stands for Pd, Ag and Cu.

**Table S1.** Rietveld refinement of XRD patterns.

| Sample | Average crystallite size (nm) | |
|---|---|---|
| | (100) direction | (002) direction |
| CdS | 16 | 36 |
| Cd(Pd)S | 13 | 31 |
| Pd/Cd(Pd)S | 16 | 40 |

**Table S2.** Fitting results of relevant standards EXAFS.

| Standard | Path | CN | Distance (Å) |
|----------|------|----|--------------|
| Pd metal | Pd–Pd | 12 (fixed) | 2.742 ± 0.001 |
| PdS | Pd–S | 4 (fixed) | 2.320 ± 0.006 |
| $K_2PdCl_4$ | Pd–Cl | 4 (fixed) | 2.310 ± 0.009 |

**Table S3.** Fitting results of catalysts EXAFS.

| Sample | Path | CN | Distance (Å) |
|---|---|---|---|
| H$_2$PdCl$_4$ in H$_2$O | Pd–O | 1.7 ± 0.5 | 2.08 ± 0.08 |
|  | Pd–Cl | 3.0 ± 0.3 | 2.292 ± 0.005 |
| H$_2$PdCl$_4$ in CH$_3$OH | Pd–O | 1.6 ± 0.6 | 2.07 ± 0.08 |
|  | Pd–Cl | 3.0 ± 0.5 | 2.280 ± 0.005 |
| Cd(Pd)S before irradiation | Pd–O | - | - |
|  | Pd–S | 3.9 ± 0.3 | 2.318 ± 0.003 |
|  | Pd–Pd | 0.5 ± 0.2 | 2.72 ± 0.02 |
| Pd/CdS before irradiation | Pd–O | 1.3 ± 0.3 | 1.99 ± 0.02 |
|  | Pd–S | 3.0 ± 0.2 | 2.308 ± 0.004 |
|  | Pd–Pd | - | - |
| Pd/Cd(Pd)S before irradiation | Pd–O | 0.8 ± 0.2 | 2.01 ± 0.02 |
|  | Pd–S | 3.3 ± 0.1 | 2.321 ± 0.003 |
|  | Pd–Pd | 0.2 ± 0.1 | 2.78 ± 0.01 |
| Cd(Pd)S | Pd–O | - | - |
|  | Pd–S | 3.3 ± 0.1 | 2.320 ± 0.003 |
|  | Pd–Pd | 0.4 ± 0.2 | 2.76 ± 0.02 |
| Pd/CdS | Pd–O | 0.8 ± 0.3 | 2.01 ± 0.03 |
|  | Pd–S | 2.0 ± 0.2 | 2.328 ± 0.007 |
|  | Pd–Pd | 0.7 ± 0.2 | 2.77 ± 0.01 |
| Pd/Cd(Pd)S | Pd–O | - | - |
|  | Pd–S | 3.1 ± 0.2 | 2.320 ± 0.004 |
|  | Pd–Pd | 0.8 ± 0.2 | 2.85 ± 0.02 |

**Table S4:** Fitting of Pd/Cd(Pd)S after irradiation in MeOH using different models comprising Pd-S, Pd-Pd and Pd-Cd scattering paths.

| Model 1* | CN | Distance (Å) | Reduced $\chi^2$ |
|---|---|---|---|
| Pd-S | 3.1 ± 0.2 | 2.320 ± 0.004 | 71.3 |
| Pd-Pd | 0.8 ± 0.2 | 2.85 ± 0.02 | |
| **Model 2** | **CN** | **Distance (Å)** | **Reduced $\chi^2$** |
| Pd-S | 3.2 ± 0.2 | 2.320 ± 0.004 | 89.5 |
| Pd-Cd | 0.7 ± 0.3 | 2.83 ± 0.02 | |
| **Model 3** | **CN** | **Distance (Å)** | **Reduced $\chi^2$** |
| Pd-S | 3.1 ± 0.2 | 2.319 ± 0.006 | |
| Pd-Pd | 1 ± 2 | 2.8 ± 0.2 | 124.3 |
| Pd-Cd | 0 ± 2 | ---- | |

* Already presented in Table S3

**Table S5.** The summary of the reported activity for photocatalytic anaerobic dehydrogenation of methanol on representative catalysts.

| Catalysts | T (ºC) | Light source | Reaction solution | $H_2$ production (mmol g$^{-1}$ h$^{-1}$) | AQY | Ref. |
|---|---|---|---|---|---|---|
| $_{0.75\%}$Cu atom-TiO$_2$ | 25 | Xe lamp (MAX-302) | 33% methanol aqueous | 16.6 | 45.5% (340nm) | 2 |
| i-Pt-TiO$_2$-150 | - | Xe lamp | 20% methanol aqueous | 17.9 | 8.6% (375 nm) | 3 |
| MoS$_2$/TiO$_2$ | - | 300 W Xe-arc lamp | 10% methanol aqueous | 2.1 | 6.4% (360 nm) | 4 |
| $_{0.5\%}$Ni/TiO$_2$ | - | UV SB-100P/F (365 nm) | 10% methanol aqueous | 31.1 | 22.2% (365 nm) | 5 |
| Ni(OH)$_2$/TiO$_2$ | - | UV-LEDs (3 W, 365 nm) | 25% methanol aqueous | 3.1 | 12.4% (365 nm) | 6 |
| $_{34\%}$Cu-TiO$_2$ | - | UV-LED | 25% methanol aqueous | 5.1 | 17.2% (365 nm) | 7 |
| $_{1.5\%}$Cu-TiO$_2$ | 40 | Xe lamp (Perfect light PLS-SXE300C) | 67% methanol aqueous | 101.7 | 56% (365 nm) | 8 |
| Pd$_{0.75}$/TiO$_2$ | 25 | 300 W Xe lamp | 25% methanol aqueous | 53.6 | 40.16% (365 nm) | 9 |
| PtCu-TiO$_2$ | 25 | 365 nm LED | 70% methanol aqueous | 241.3 | 56.2% (365 nm) | 10 |
| PtCu-TiO$_2$ | 40 | 365 nm LED | 70% methanol aqueous | 262.9 | 73.4% (365 nm) | 10 |
| PtCu-TiO$_2$ | 50 | 365 nm LED | 70% methanol aqueous | 320.0 | 83.1% (365 nm) | 10 |
| PtCu-TiO$_2$ | 60 | 365 nm LED | 70% methanol aqueous | 383.0 | 96.6% (365 nm) | 10 |
| PtCu-TiO$_2$ | 70 | 365 nm LED | 70% methanol aqueous | 476.8 | 99.2% (365 nm) | 10 |
| Ni/CdS | 20 | 300 W Xe lamp | 4 mM NiCl$_2$/methanol solution | 48.2 | 95% (405 nm) | 11 |
| Ni/CdS | 20 | 447 nm LED | anhydrous methanol | 7.5 | 38% (447 nm) | 12 |
| TiO$_2$/Cu_50 | 25 | 300 W Xe lamp | anhydrous methanol | 17.8 | 16.4% (365 nm) | 13 |
| Pd/Cd(Pd)S | 25 | 452 nm LED | anhydrous methanol | 208.4 | 87% (452 nm) | This work |

## Supporting References